\documentclass[english,aip,journal=jcp,reprint]{revtex4-1}
\usepackage[T1]{fontenc}
\usepackage[utf8]{inputenc}
\usepackage{refstyle}
\usepackage{graphicx}
\usepackage[version=4]{mhchem}
\usepackage{microtype}
\usepackage[unicode=true,pdfusetitle,hyperindex,
 bookmarks=true,bookmarksnumbered=false,bookmarksopen=false,
 breaklinks=true,pdfborder={0 0 1},backref=false,colorlinks=false]
 {hyperref}

\makeatletter

\AtBeginDocument{\providecommand\eqref[1]{\ref{eq:#1}}}
\AtBeginDocument{\providecommand\figref[1]{\ref{fig:#1}}}
\AtBeginDocument{\providecommand\tabref[1]{\ref{tab:#1}}}
\providecommand{\tabularnewline}{\\}
\RS@ifundefined{subsecref}
  {\newref{subsec}{name = \RSsectxt}}
  {}
\RS@ifundefined{thmref}
  {\def\RSthmtxt{theorem~}\newref{thm}{name = \RSthmtxt}}
  {}
\RS@ifundefined{lemref}
  {\def\RSlemtxt{lemma~}\newref{lem}{name = \RSlemtxt}}
  {}

\makeatother

\begin{document}

\setcounter{topnumber}{3}
\setcounter{bottomnumber}{3}
\setcounter{totalnumber}{6}

\title{Efficient implementation of the superposition of atomic potentials
initial guess for electronic structure calculations in Gaussian basis sets}
\author{Susi Lehtola}
\email{susi.lehtola@alumni.helsinki.fi}
\affiliation{Department of Chemistry, University of Helsinki, P.O. Box 55 (A. I. Virtasen
  aukio 1), FI-00014 Helsinki, Finland.}
\author{Lucas Visscher}
\email{l.visscher@vu.nl}
\affiliation{Division of Theoretical Chemistry, Vrije Universiteit Amsterdam,De
  Boelelaan 1083, 1081 HV Amsterdam, The Netherlands.}
\author{Eberhard Engel}
\affiliation{Center for Scientific Computing, J. W. Goethe-Universität Frankfurt,
Max-von-Laue-Strasse 1, D-60438 Frankfurt am Main, Germany.}
\begin{abstract}
The superposition of atomic potentials (SAP) approach has recently
been shown to be a simple and efficient way to initialize electronic
structure calculations {[}S. Lehtola, J. Chem. Theory Comput. 15, 1593
  (2019){]}. Here, we study the differences between effective
potentials from fully numerical density functional and optimized
effective potential calculations for fixed configurations. We find
that the differences are small, overall, and choose exchange-only
potentials at the local density approximation level of theory computed
on top of Hartree--Fock densities as a good compromise. The
differences between potentials arising from different atomic
configurations are also found to be small at this level of theory.

Furthermore, we discuss the efficient Gaussian-basis implementation
of SAP via error function fits to fully numerical atomic radial potentials.
The guess obtained from the fitted potentials can be easily implemented
in any Gaussian-basis quantum chemistry code in terms of two-electron
integrals. Fits covering the whole periodic table from H to Og are
reported for non-relativistic as well as fully relativistic four-component
calculations that have been carried out with fully numerical approaches.
\end{abstract}
\maketitle
\global\long\def\ERI#1#2{(#1|#2)}%
\global\long\def\bra#1{\Bra{#1}}%
\global\long\def\ket#1{\Ket{#1}}%
\global\long\def\erf#1{\text{\,erf\,}(#1)}%
\global\long\def\erfc#1{\text{\,erfc\,}(#1)}%

\newcommand*\ie{\emph{i.e.}}
\newcommand*\eg{\emph{e.g.}}
\newcommand*\citeref[1]{ref. \citenum{#1}}
\newcommand*\citerefs[1]{refs. \citenum{#1}}
\newcommand*\HelFEM{{\sc HelFEM}}
\newcommand*\Erkale{{\sc Erkale}}
\newcommand*\Gaussian{{\sc Gaussian}}
\newcommand*\Grasp{{\sc Grasp}}
\newcommand*\Dirac{{\sc Dirac}}
\newcommand*\PsiFour{{\sc Psi4}}
\newcommand*\Mathematica{{\sc Mathematica}}

\section{Introduction \label{sec:intro}}

In order to perform an electronic structure calculation, an initial
guess is necessary for the one-particle states \ie{} orbitals, and
several types of guesses have been proposed over the
years.\cite{Lehtola2019} The focus of the present work is the
superposition of atomic potentials (SAP), which is arguably a very old
idea with roots dating back at least to the late
1960s.\citep{Green1969, Whalen1972} However, SAP was apparently
forgotten for a long time, assumedly due to issues that were only
recently fully resolved.\cite{Lehtola2019}

But, if one is interested only in Gaussian-basis calculations, then
adopting an atomic potential arising from a spherically symmetric
Gaussian expansion of a fictitious electron density (yielding error
functions as is well-known) leads to facile evaluation of the
necessary matrix elements. The formation of the guess reduces to the
same two-electron integrals that are used in the subsequent
self-consistent field (SCF) procedure, which thus already exist in all
Gaussian-basis quantum chemistry programs. Such special variants of
the SAP guess were proposed by Whitten and
coworkers.\citep{Nazari2017, Whitten2019, Whitten2019a} They developed
potentials derived from Gaussian pseudo-electron densities, which were
optimized for specific elements described with a specific Gaussian
orbital basis set, and embedded in specific chemical
environments.\citep{Nazari2017, Whitten2019, Whitten2019a}

In contrast to the Gaussian-basis approach pursued by Whitten and
coworkers, where the potentials are tailored to specific chemical
environments, a parameter-free variant of the SAP guess based on
potentials that are determined from fully numerical\cite{Lehtola2019c}
atomic density functional calculations at the complete basis set limit
was presented in \citeref{Lehtola2019}. The resulting SAP guess proved
to be the most accurate out of the seven types of initial guesses
considered in \citeref{Lehtola2019}, judged by the projection of the
guess orbitals onto the SCF solution. An improved method for
determining the atomic potentials necessary for the procedure has been
recently presented in \citeref{Lehtola2019e}. A straightforward
extension of the work in \citerefs{Lehtola2019} and
\citenum{Nazari2017} has also been recently suggested in
\citeref{Laikov2020}, in which a universal atomic potential is
employed as in \citeref{Lehtola2019}, but instead of real-space
calculations at the basis set limit the potentials are obtained for a
small Gaussian basis set and biased for molecular calculations on the
lines of \citeref{Nazari2017}. However, the optimization in
\citeref{Laikov2020} was restricted to fixing the wrong asymptotic
behavior of the optimized effective potential discussed by one of the
present authors in \citeref{Lehtola2019-v1preprint}, and instead of
minimizing the resulting guess energy as Whitten and coworkers in
\citerefs{Nazari2017, Whitten2019, Whitten2019a}, the procedure of
\citeref{Laikov2020} maximizes the overlap of the guess orbitals onto
the SCF solution according to the procedure first introduced in
\citeref{Lehtola2019}.

All variants of SAP (including \citerefs{Green1969, Whalen1972,
  Nazari2017, Whitten2019, Whitten2019a, Lehtola2019} and
\citenum{Laikov2020}) assume that the potential the electrons feel in
a molecule can be accurately modeled by a simple sum of atomic
potentials.  Although potentials optimized for molecular calculations
may have benefits, the optimization makes them less general than ones
derived from atomic calculations. The transferability of optimized
potentials across basis sets is not inherently clear, as the
optimizations are typically carried out in small basis sets, and any
possible artefacts of the small-basis optimization may only become
visible in calculations with extended basis sets, or in applications
to large molecules. In contrast, potentials derived strictly from
first principles are appealing as they can be routinely obtained at
the complete basis set limit, thus guaranteeing transferability
between basis sets and large systems.\cite{Lehtola2019} They can also
be customized for a specific purpose, if so desired. For instance,
spin-polarized orbitals for e.g. antiferromagnetically coupled systems
can be obtained straightforwardly by employing alternating potentials
on the atoms. If one employs the potential corresponding to the atomic
majority spin channel from a numerical atomic structure calculation,
the resulting guess orbital will place more density on the atom than
if one uses the minority spin potential which is less attractive due
to a smaller amount of exchange.

To facilitate the implementation of the first-principles SAP guess
described in \citeref{Lehtola2019} in Gaussian-basis quantum chemistry
programs, in the present work we report error function expansions of
atomic effective potentials derived from fully numerical atomic
calculations. With these fits, the SAP guess can be implemented in any
Gaussian-basis quantum chemistry program in terms of three-center
two-electron integrals that are familiar from
resolution-of-the-identity methods.\cite{Vahtras1993} We would
especially like to point out that the implementation of the fitted SAP
guess is fully analogous to the computation of the nuclear attraction
matrix elements for finite nuclei with Gaussian
distributions\citep{Visscher1997, Andrae2000} that is already
available in several program packages. An implementation of SAP based
on this technique has been available in the \Dirac{} program since its
2016 release, but it has not yet been described.

The outline of the work is as follows. We will briefly summarize the
SAP method in \secref{method}. Various parameters of the calculations
and the fitting procedures are detailed in
\secref{compdet}. \Secref{results} presents fits to both
non-relativistic and fully relativistic four-component calculations,
which have been obtained with the \HelFEM{} and \Grasp{} programs,
respectively. Molecular applications of the potentials are shown in
\secref{molecular}. The article concludes with a brief summary and
discussion in \secref{summary}. Atomic units are used throughout the
text.

\section{Method \label{sec:method}}

\subsection{Superposition of atomic potentials \label{sec:sap}}

As the name suggests, the basic idea in the SAP approach is to obtain
approximate molecular orbitals from an effective one-particle
Hamiltonian (shown here in the non-relativistic case for simplicity)
\begin{equation}
\hat{H}=-\frac{1}{2}\nabla^{2}+V^{\text{SAP}}(\boldsymbol{r})=-\frac{1}{2}\nabla^{2}-\sum_{A}\frac{Z_{A}(r_{A})}{r_{A}}\label{eq:Hsap}
\end{equation}
where the effective potential $V^{\text{SAP}}(\boldsymbol{r})$ is
obtained as a superposition of atomic potentials. The atomic
potentials can in turn be rewritten in terms of effective nuclear
charges $Z_{A}$, as seen at a distance
$r_{A}=\left|\boldsymbol{r}-\boldsymbol{R}_{A}\right|$ away from the
nucleus $A$ at $\boldsymbol{R}_{A}$. As the potentials $V^\text{SAP}$
are to be local, we will define the exchange-(correlation) part of the
potential in terms of density functional approximations (DFA) to
density functional theory\cite{Hohenberg1964, Kohn1965} (DFT) as in
\citeref{Lehtola2019}.

Due to the radial dependence of the effective charge, the reliable
calculation of the matrix elements of the Hamiltonian in \eqref{Hsap}
appears difficult. However, the realization made in
\citeref{Lehtola2019} was that similar numerical problems also appear
in density functional approaches; for example, the multicenter
quadrature scheme of \citeref{Becke1988} is a suitable solution in the
case of atomic basis set calculations. The matrix elements of
\eqref{Hsap} can be computed by minor modifications to existing
density functional routines; such implementations are now available in
\Erkale{},\citep{erkale, Lehtola2012} \PsiFour{},\citep{Parrish2017}
and the fully numerical \HelFEM{} program.\citep{HelFEM, Lehtola2019a,
  Lehtola2019b} (A similar approach has also
been used in the calculation of matrix elements for the ``maximum of
atomic potentials'' approach to the zeroth-order regular expansion
approximation\cite{Lenthe1993} within the ADF program.\cite{ADF})

In contrast, the implementation in \Dirac{}, like the work in
\citerefs{Nazari2017, Whitten2019, Whitten2019a} and
\citenum{Laikov2020}, is based on expanding the electronic part of the
radial potential in terms of the potentials of primitive normalized
$s$-type Gaussians $g_p(r) = (\alpha_p/\pi)^{3/2} \exp(-\alpha_p r^2)$
that have the simple expression\cite{Boys1950} $V_p (r) =
\erf{\sqrt{\alpha_p}r} / r$, where erf is the error function.  The
comparison of \eqref{Hsap} to the above shows that this amounts simply
to expanding $Z^\text{el}(r)$ in a set of error functions
$\phi^0_p=\erf{\beta_pr}$ with arguments $\beta_p=\sqrt{\alpha_p}$ and
expansion coefficients $\{c_p\}_{p=1}^N$. The matrix elements of the
fitted potential in an atom-centered orbital basis $\{\chi_i\}$
\begin{equation}
V^\text{el}_{ij} = \langle i|V|j \rangle=\int\chi_{i}(\boldsymbol{r})V(r_{A})\chi_{j}(\boldsymbol{r}){\rm d}^{3}r\label{eq:mat-el}
\end{equation}
become easy to evaluate, as rewriting the error functions in integral
form as potentials arising from the normalized Gaussian functions
$|\alpha_p)$ yields an expression in terms of three-center
two-electron integrals (3C-TEIs)
\begin{equation}
V^\text{el}_{ij} = -\sum_{p=1}^{N} c_{p}(ij|\alpha_{p}), \label{eq:sap-fitpot}
\end{equation}
where the negative sign comes from \eqref{Hsap}. 3C-TEIs are familiar
from resolution-of-the-identity methods.\cite{Vahtras1993}
\Eqref{sap-fitpot} also bears a strong similarity to the description
of finite nuclei with Gaussian distributions in quantum chemical
calculations,\citep{Visscher1997, Andrae2000} where only a single
auxiliary function $\alpha$ is used; its exponent being determined by
the size of the nucleus and its coefficient coinciding with the
nuclear charge.

3C-TEIs can be obtained as a special case of general two-electron
integrals---the basic ingredient of Gaussian-basis quantum chemistry
programs---that can be evaluated analytically in an efficient
fashion;\cite{McMurchie1978, Obara1986} alternatively, 3C-TEIs can be
evaluated even more efficiently with specialized
approaches.\cite{Ahlrichs2004} The (approximate) fit of the radial
potential thereby allows one to circumvent the need for (approximate)
quadratures of $V(r)$ pursued in \citeref{Lehtola2019}.

\subsection{Fitting scheme \label{sec:fitscheme}}

The fitting error with fitting coefficients $\{c_p\}$ and a fitting
basis $\{\phi^0_p\}$ is given by
\begin{equation}
\tau=\int_0^\infty \left[Z^\text{el} (r)-\sum_{p}c_{p}\phi^0_{p}(r)\right]^{2}{\rm d}r.\label{eq:tau}
\end{equation}
Although error functions were adopted as the fitting basis in
\secref{sap}, a complication with this choice is that the overlap
matrix
\begin{equation}
  S^0_{pq} = \int_0^\infty \phi^0_{p}(r) \phi^0_{q}(r) {\rm d}r \label{eq:ovl}
\end{equation}
is divergent. In order to determine the coefficients, it is therefore
better to rewrite the problem in terms of complementary error
functions. Taking $\phi_{p}(r) = \erfc{\beta_p r}$ leads to an
analytical expression for the overlap matrix
\begin{equation}
  S_{pq} = \frac{\beta_{p}+\beta_{q}-\sqrt{\beta_{p}^{2}+\beta_{q}^{2}}}{\beta_{p}\beta_{q}\sqrt{\pi}},\label{eq:erfc-overlap}
\end{equation}
meaning that such a basis is well-behaved for an expansion. All that
remains is to rewrite the original fitting problem in terms of
erfc's. As $\phi^0_p(r) = \erf{\beta_p r} = 1-\erfc{\beta_p r} = 1 -
\phi_p(r)$, \eqref{tau} can be rewritten as
\begin{equation}
\tau=\int_0^\infty \left[ \left(Z^\text{el} (r)-\sum_{p}c_{p} \right)  + \sum_{p} c_{p} \phi_{p}(r)\right]^{2}{\rm d}r \label{eq:tauc}
\end{equation}
without changing the meaning of the coefficients $c_p$.

The DFA potentials used in the present work have the important
property that far away the effective charge goes to
zero,\cite{Lehtola2019} $Z_A(\infty) =Z_A - Z_A^\text{el}(\infty) = 0$. Imposing this long-range
limit translates into the condition
\begin{equation}
\sum_{p} c_{p}=-Z \label{eq:fit-cond}
\end{equation}
where $Z$ is the nuclear charge. (Note that atomic potentials that do
not satisfy this requirement lead to molecular potentials that become
worse and worse in increasing system size, see the discussion in
\citeref{Lehtola2019-v1preprint}.)

Furthermore, we can rewrite \eqref{tauc} in terms of an effective
charge as $Z(r) = Z - Z^\text{el}(r)$ by using \eqref{fit-cond}. This
redefinition of $Z(r)$ coincides with the screened nuclear potential
in the case of a point nucleus, but this mathematical trick works
equally well with a finite nucleus: the important thing to notice here
is that the nuclear model does not enter into the fits of the
potential generated by the electrons. \Eqref{tauc} thus becomes
\begin{equation}
\tau=\int_0^\infty \left[ Z(r) - \sum_{p} c_{p} \phi_{p}(r)\right]^{2}{\rm d}r.\label{eq:taucnew}
\end{equation}

Although \eqref{taucnew} already allows fits to the redefined $Z(r)$,
these fits may still violate the charge neutrality condition,
\eqref{fit-cond}. The condition can be enforced by treating the
coefficient of the steepest function as a dependent variable
\begin{equation}
 c_{n} = Z - \sum_{p=1}^{n-1} c_{p}, \label{eq:c-cond}
\end{equation}
so that the fitting problem for the $n-1$ remaining coefficients
becomes
\begin{align}
  \tau & =\int_0^\infty \left[ \left[Z(r) - Z \phi_{n}(r) \right] - \sum_{p=1}^{n-1} c_{p} [\phi_{p}(r) -\phi_{n}(r)] \right]^{2}{\rm d}r \nonumber \\
  & =\int_0^\infty \left[ \tilde{Z}(r) - \sum_{p=1}^{n-1} c_{p} \tilde{\phi}_{p}(r) \right]^{2}{\rm d}r. \label{eq:taufinal}
\end{align}
The error is minimized by coefficients ${\bf c}$ that satisfy
\begin{equation}
\frac{\partial\tau}{\partial c_{q}}=-2\int_0^\infty \left[\tilde{Z}(r)-\sum_{p}c_{p}\tilde{\phi}_{p}(r)\right]\tilde{\phi}_{q}(r){\rm d}r=0\label{eq:dtaudc}
\end{equation}
from which
\begin{equation}
  \tilde{\bf S} {\bf c}=\tilde{\bf Z}, \label{eq:cequation}
\end{equation}
where
\begin{align}
  \tilde{Z}_{q} & =\int_0^\infty \tilde{Z}(r)\tilde{\phi}_{q}(r){\rm d}r \label{eq:Zproj} \\
  \tilde{S}_{pq} & =\int_0^\infty \tilde{\phi}_{p}(r)\tilde{\phi}_{q}(r){\rm d}r. \label{eq:Sproj}
\end{align}
\Eqref{cequation} can be solved for the $n-1$ coefficients, \eg{}, by
computing the inverse overlap matrix via the canonical
orthogonalization procedure\cite{Lowdin1956} in which eigenvectors
with eigenvalues smaller than $10^{-7}$ are omitted, after which the
dependent coefficient is calculated from \eqref{c-cond}. The basis set
is normalized before the canonical orthogonalization procedure to
ensure proper conditioning of the eigenproblem.\cite{Lehtola2020}
While one may in principle define any coefficient as the dependent
coefficient, eliminating the coefficient of the tightest function has
the advantage that the function $\tilde{Z}(r) = Z(r)-Z\phi_n (r)$
exhibits the fastest decay to zero for $r \to 0$.

The function $Z(r)$ from \HelFEM{} or \Grasp{} is essentially exact:
it yields the energy of the atom at the complete basis set limit.
Also the matrix elements of \eqref{Zproj} can be evaluated exactly,
\ie{} without any significant error, with the numerical grid from
\HelFEM{} or \Grasp{}, since $Z(r)$ is accurately known where it is
non-zero. The overlap integrals of \eqref{Sproj}, in turn, are
evaluated analytically via \eqref{erfc-overlap}. The only potentially
significant source of numerical errors in the fitting procedure is the
calculation of the total fit error $\tau$ in \eqref{tau}, as this
quantity may not be evaluated accurately by quadrature on the
grid.

Especially, the grid is deficient in regions where $Z(r)=0$, that is,
far away from the nucleus: although the fit coefficients are
accurately evaluated, the resulting fit error is not. For instance, if
the most diffuse fitting functions are nonzero at the practical
infinity $r_{\infty}$ of the fully numerical calculation,
$\erfc{\beta_\text{min} r_{\infty}} \neq 0$, the fit error in the
potential from $r=r_\infty$ to $r=\infty$ is completely neglected by
the quadrature. A more accurate evaluation of the fitting error $\tau$
can be achieved by the addition of a penalty term
\begin{equation}
  \tau \to \tau + \sum_{pq} c_{p} [S_{pq} - \tilde{S}_{pq}]
  c_{q} \label{eq:penalty}
\end{equation}
where $\tilde{\bf S}$ is the quadrature evaluation of the overlap
matrix. Fit functions that are accurately described on the grid carry
no penalty as $\tilde{S}_{pq} \approx S_{pq}$. Otherwise, the fit
functions pick up a penalty, as they should: if the functions are not
accurately described on the grid, their form also ill describes
$Z(r)$, whose grid representation is known to be exact. This means
that in addition to describing the quadrature error in $(r_\infty,
\infty)$ discussed above, the term may also describe quadrature errors
in $(0,r_\infty)$.

The linear expansion coefficients $c_{p}$ are unambiguously determined
by \eqrangeref{cequation}{Sproj} once the primitives $\beta_{p}$ have
been chosen. For simplicity, we use a universal set of even-tempered
parameters $\beta_p = \beta_0 \gamma^p$ where $\beta_0$ and $\gamma$
are constants and $p$ are integers, as such expansions afford an easy
way to approach the complete (fitting) basis set
limit.\cite{Feller1979} The actual procedure for the formation of the
fitting basis follows the procedure of \citeref{Lehtola2020b}. First,
the best single $\beta_p$ parameter is found (also allowing negative
values of $p$), after which steeper and more diffuse functions are
added into the fitting basis set one by one until the complete
fitting-basis-set limit has been achieved, defined as the point at
which the fit error only goes up when further functions are added due
to finite numerical accuracy. Next, because the fit error often
plateaus long before the minimum error for the given $\beta_0$ and
$\gamma$ is found, the shortest expansion that yields an error within
5\% of the minimum is chosen for production purposes.

However, this set of fits that yields (close to) the lowest possible
error for each element with given $\beta_0$ and $\gamma$ is still
suboptimal, as fixed values for $\beta_0$ and $\gamma$ afford fits of
a different quality for different elements in the periodic table. The
fit error $\tau$ can be made especially small for the lightest
elements, while the error tends to increase with $Z$. A balanced fit
has a uniform accuracy across $Z$; this is achieved by truncating the
fits further so that $\tau'(Z) \leq \max_{Z} \tau(Z)$ still holds,
that is, so that the fit error of the truncated fits $\tau'(Z)$ is
bound by the largest error of the original fits $\tau(Z)$. The
truncation results in a major compactification of the tabulated fits
by reducing the fits for the light elements to a fraction of their
original size.

\section{Computational Details \label{sec:compdet}}

Non-relativistic calculations were performed with
\HelFEM{}\cite{HelFEM, Lehtola2019a, Lehtola2019e} using 20 radial
elements and a value of the practical infinity $r_{\infty}=40a_{0}$;
the resulting \HelFEM{} energies are converged beyond nanohartree
accuracy for all atoms.\cite{Lehtola2019a, Lehtola2019e} The \HelFEM{}
calculations employed fractional occupations resulting in spherically
symmetric densities, and the ground state for each element was found
automatically by a brute force search,\citep{Lehtola2019e} leading
\eg{} to the unrestricted Hartree--Fock (HF) configurations shown in
\tabref{conf}. The equations for calculating the radial potential in
the finite element formalism used in \HelFEM{} have been presented in
\citeref{Lehtola2019e} to which we refer for further details.

\begin{table*}
\begin{tabular}{lllllrr|lllllrr|lllllrr}
 & $ n_s $ & $ n_p $ & $ n_d $ & $ n_f $ & $ E/E_h $ & $ M $ &  & $ n_s $ & $ n_p $ & $ n_d $ & $ n_f $ & $ E/E_h $ & $ M $ &  & $ n_s $ & $ n_p $ & $ n_d $ & $ n_f $ & $ E/E_h $ & $ M $\tabularnewline 
\hline 
\hline 
H & 1 & 0 & 0 & 0 & $ -0.500000 $ & 2 & Nb & 8 & 18 & 15 & 0 & $ -3753.558738 $ & 6 & Tl & 12 & 25 & 30 & 14 & $ -18961.760416 $ & 2\tabularnewline
He & 2 & 0 & 0 & 0 & $ -2.861680 $ & 1 & Mo & 9 & 18 & 15 & 0 & $ -3975.552980 $ & 7 & Pb & 11 & 27 & 30 & 14 & $ -19523.935177 $ & 5\tabularnewline
Li & 3 & 0 & 0 & 0 & $ -7.432751 $ & 2 & Tc & 10 & 18 & 15 & 0 & $ -4204.794932 $ & 6 & Bi & 12 & 27 & 30 & 14 & $ -20095.588624 $ & 4\tabularnewline
Be & 4 & 0 & 0 & 0 & $ -14.573023 $ & 1 & Ru & 10 & 18 & 16 & 0 & $ -4441.293960 $ & 5 & Po & 12 & 28 & 30 & 14 & $ -20676.415929 $ & 3\tabularnewline
B & 4 & 1 & 0 & 0 & $ -24.415026 $ & 2 & Rh & 8 & 18 & 19 & 0 & $ -4685.642225 $ & 2 & At & 12 & 29 & 30 & 14 & $ -21266.785190 $ & 2\tabularnewline
C & 3 & 3 & 0 & 0 & $ -37.599255 $ & 5 & Pd & 8 & 18 & 20 & 0 & $ -4937.921024 $ & 1 & Rn & 12 & 30 & 30 & 14 & $ -21866.772241 $ & 1\tabularnewline
N & 4 & 3 & 0 & 0 & $ -54.404548 $ & 4 & Ag & 9 & 18 & 20 & 0 & $ -5197.698943 $ & 2 & Fr & 13 & 30 & 30 & 14 & $ -22475.858834 $ & 2\tabularnewline
O & 4 & 4 & 0 & 0 & $ -74.622399 $ & 3 & Cd & 10 & 18 & 20 & 0 & $ -5465.133143 $ & 1 & Ra & 14 & 30 & 30 & 14 & $ -23094.303666 $ & 1\tabularnewline
F & 4 & 5 & 0 & 0 & $ -99.164711 $ & 2 & In & 10 & 19 & 20 & 0 & $ -5740.102296 $ & 2 & Ac & 14 & 30 & 31 & 14 & $ -23722.088791 $ & 2\tabularnewline
Ne & 4 & 6 & 0 & 0 & $ -128.547098 $ & 1 & Sn & 9 & 21 & 20 & 0 & $ -6022.853866 $ & 5 & Th & 12 & 30 & 34 & 14 & $ -24359.586764 $ & 5\tabularnewline
Na & 5 & 6 & 0 & 0 & $ -161.858954 $ & 2 & Sb & 10 & 21 & 20 & 0 & $ -6313.487048 $ & 4 & Pa & 13 & 30 & 30 & 18 & $ -25006.654223 $ & 6\tabularnewline
Mg & 6 & 6 & 0 & 0 & $ -199.614636 $ & 1 & Te & 10 & 22 & 20 & 0 & $ -6611.692943 $ & 3 & U & 12 & 30 & 30 & 20 & $ -25664.034255 $ & 7\tabularnewline
Al & 6 & 7 & 0 & 0 & $ -241.803440 $ & 2 & I & 10 & 23 & 20 & 0 & $ -6917.876506 $ & 2 & Np & 12 & 30 & 30 & 21 & $ -26331.520612 $ & 8\tabularnewline
Si & 5 & 9 & 0 & 0 & $ -288.763297 $ & 5 & Xe & 10 & 24 & 20 & 0 & $ -7232.138364 $ & 1 & Pu & 13 & 30 & 30 & 21 & $ -27008.844714 $ & 9\tabularnewline
P & 6 & 9 & 0 & 0 & $ -340.719275 $ & 4 & Cs & 11 & 24 & 20 & 0 & $ -7553.933772 $ & 2 & Am & 14 & 30 & 30 & 21 & $ -27695.900612 $ & 8\tabularnewline
S & 6 & 10 & 0 & 0 & $ -397.386801 $ & 3 & Ba & 12 & 24 & 20 & 0 & $ -7883.543827 $ & 1 & Cm & 14 & 30 & 31 & 21 & $ -28392.659412 $ & 9\tabularnewline
Cl & 6 & 11 & 0 & 0 & $ -459.339556 $ & 2 & La & 12 & 24 & 21 & 0 & $ -8220.952378 $ & 2 & Bk & 14 & 30 & 30 & 23 & $ -29099.513303 $ & 6\tabularnewline
Ar & 6 & 12 & 0 & 0 & $ -526.817513 $ & 1 & Ce & 10 & 24 & 24 & 0 & $ -8566.569397 $ & 5 & Cf & 13 & 30 & 30 & 25 & $ -29816.800544 $ & 5\tabularnewline
K & 7 & 12 & 0 & 0 & $ -599.164870 $ & 2 & Pr & 12 & 24 & 20 & 3 & $ -8920.395478 $ & 4 & Es & 12 & 30 & 30 & 27 & $ -30544.612534 $ & 2\tabularnewline
Ca & 8 & 12 & 0 & 0 & $ -676.758186 $ & 1 & Nd & 11 & 24 & 20 & 5 & $ -9283.115285 $ & 7 & Fm & 12 & 30 & 30 & 28 & $ -31282.870930 $ & 1\tabularnewline
Sc & 8 & 13 & 0 & 0 & $ -759.574264 $ & 2 & Pm & 10 & 24 & 20 & 7 & $ -9654.657927 $ & 8 & Md & 13 & 30 & 30 & 28 & $ -32031.172211 $ & 2\tabularnewline
Ti & 7 & 12 & 3 & 0 & $ -848.081411 $ & 5 & Sm & 11 & 24 & 20 & 7 & $ -10034.895222 $ & 9 & No & 14 & 30 & 30 & 28 & $ -32789.512140 $ & 1\tabularnewline
V & 6 & 12 & 5 & 0 & $ -942.764789 $ & 6 & Eu & 12 & 24 & 20 & 7 & $ -10423.550567 $ & 8 & Lr & \emph{14} & \emph{31} & \emph{30} & \emph{28} & \emph{$ -33557.817804 $} & \emph{2}\tabularnewline
Cr & 7 & 12 & 5 & 0 & $ -1043.356782 $ & 7 & Gd & 12 & 24 & 21 & 7 & $ -10820.539254 $ & 9 & Rf & 12 & 30 & 34 & 28 & $ -34336.517013 $ & 5\tabularnewline
Mn & 8 & 12 & 5 & 0 & $ -1149.869841 $ & 6 & Tb & 10 & 24 & 24 & 7 & $ -11226.259522 $ & 12 & Db & 12 & 30 & 35 & 28 & $ -35125.642045 $ & 6\tabularnewline
Fe & 8 & 13 & 5 & 0 & $ -1262.258941 $ & 7 & Dy & 12 & 24 & 20 & 10 & $ -11640.492034 $ & 5 & Sg & 13 & 30 & 35 & 28 & $ -35924.911091 $ & 7\tabularnewline
Co & 8 & 12 & 7 & 0 & $ -1380.935491 $ & 4 & Ho & 11 & 24 & 20 & 12 & $ -12064.271314 $ & 4 & Bh & 14 & 30 & 35 & 28 & $ -36734.336697 $ & 6\tabularnewline
Ni & 6 & 12 & 10 & 0 & $ -1506.669759 $ & 1 & Er & 10 & 24 & 20 & 14 & $ -12497.495591 $ & 1 & Hs & 12 & 30 & 38 & 28 & $ -37553.987931 $ & 3\tabularnewline
Cu & 7 & 12 & 10 & 0 & $ -1638.964246 $ & 2 & Tm & 11 & 24 & 20 & 14 & $ -12940.015784 $ & 2 & Mt & 12 & 30 & 39 & 28 & $ -38384.346407 $ & 2\tabularnewline
Zn & 8 & 12 & 10 & 0 & $ -1777.848116 $ & 1 & Yb & 12 & 24 & 20 & 14 & $ -13391.456193 $ & 1 & Ds & 12 & 30 & 40 & 28 & $ -39225.264332 $ & 1\tabularnewline
Ga & 8 & 13 & 10 & 0 & $ -1923.187642 $ & 2 & Lu & 12 & 25 & 20 & 14 & $ -13851.703406 $ & 2 & Rg & 13 & 30 & 40 & 28 & $ -40076.354892 $ & 2\tabularnewline
Ge & 8 & 14 & 10 & 0 & $ -2075.267721 $ & 3 & Hf & 10 & 24 & 24 & 14 & $ -14321.061494 $ & 5 & Cn & 14 & 30 & 40 & 28 & $ -40937.797856 $ & 1\tabularnewline
As & 8 & 15 & 10 & 0 & $ -2234.239855 $ & 4 & Ta & 10 & 24 & 25 & 14 & $ -14799.827000 $ & 6 & Nh & 14 & 31 & 40 & 28 & $ -41809.475125 $ & 2\tabularnewline
Se & 8 & 16 & 10 & 0 & $ -2399.761455 $ & 3 & W & 11 & 24 & 25 & 14 & $ -15287.662265 $ & 7 & Fl & 13 & 33 & 40 & 28 & $ -42691.604385 $ & 5\tabularnewline
Br & 8 & 17 & 10 & 0 & $ -2572.317356 $ & 2 & Re & 12 & 24 & 25 & 14 & $ -15784.544119 $ & 6 & Mc & 14 & 33 & 40 & 28 & $ -43584.201351 $ & 4\tabularnewline
Kr & 8 & 18 & 10 & 0 & $ -2752.054977 $ & 1 & Os & 12 & 24 & 26 & 14 & $ -16290.475039 $ & 5 & Lv & 14 & 34 & 40 & 28 & $ -44487.022704 $ & 3\tabularnewline
Rb & 9 & 18 & 10 & 0 & $ -2938.357567 $ & 2 & Ir & 10 & 24 & 29 & 14 & $ -16806.002785 $ & 2 & Ts & 14 & 35 & 40 & 28 & $ -45400.387450 $ & 2\tabularnewline
Sr & 10 & 18 & 10 & 0 & $ -3131.545686 $ & 1 & Pt & 10 & 24 & 30 & 14 & $ -17331.121868 $ & 1 & Og & 14 & 36 & 40 & 28 & $ -46324.355815 $ & 1\tabularnewline
Y & 10 & 19 & 10 & 0 & $ -3331.575414 $ & 2 & Au & 11 & 24 & 30 & 14 & $ -17865.400624 $ & 2 &  & & & & & & \tabularnewline
Zr & 8 & 18 & 14 & 0 & $ -3538.801672 $ & 5 & Hg & 12 & 24 & 30 & 14 & $ -18408.991495 $ & 1 &  & & & & & & \tabularnewline
\end{tabular}
\caption{Non-relativistic spin-unrestricted spherical Hartree--Fock
  configurations used for the \HelFEM{} calculations. Legend: spin
  multiplicity $ M $, number of $ s $ electrons $ n_s $, number of $ p
  $ electrons $ n_p $, number of $ d $ electrons $ n_d $, number of $
  f $ electrons $ n_f $, total energy $ E $. A lower configuration for
  Lr was found in the brute-force search but it failed to converge,
  due to which a low-lying excited state is used
  instead. \label{tab:conf}}
\end{table*}

The relativistic calculations were carried out with a modified
version\cite{Graspdft} of \Grasp{}\cite{Dyall1989} using the settings
described in \citeref{Visscher1997}; in short, the Gaussian nuclear
model\cite{Visscher1997} was used in combination with the average level
(AL) option of GRASP to provide a balanced description of the valence
levels. The resulting Hartree--Fock total and orbital energies have
been presented in \citeref{Visscher1997}.

In order to study the accuracy of the density functional potentials,
optimized effective potential (OEP) calculations were performed as
detailed in \citeref{Engel1993}. Both the non-relativistic radial
Kohn--Sham equations and the OEP integral equation for the exact
exchange potential were solved fully numerically on a logarithmic
radial grid containing 4000 points to obtain high accuracy.  The
Krieger--Li--Iafrate identity\cite{Krieger1990a} was applied for the
normalization of the exchange potential. In the case of open
spin-subshells, the spin-up and spin-down Kohn--Sham potentials were
averaged.

\section{Results \label{sec:results}}

\subsection{Form of the potential \label{sec:potform}}

In order to select the form of the radial potential, we study the Fe
atom in its $4s^{2}3d^{6}$ quintet ground state. Because the state
should exhibiting significant spin-polarization, it should serve well
to illustrate differences in the possible choices for the radial
potential. We will restrict the present study to exchange-only
calculations, since the best results in \citeref{Lehtola2019} were
obtained with exchange-only potentials.

Four kinds of calculations were performed with \HelFEM{} with
fractional occupations, using the methodology presented in
\citeref{Lehtola2019e}. The first three are fully self-consistent
calculations with exchange either described by the local density
approximation (LDA),\citep{Bloch1929, Dirac1930} the
PBE\citep{Perdew1996} functional, or the EV93\citep{Engel1993a}
functional; PBE and EV93 are both generalized gradient functionals.
The fourth is a hybrid procedure in which the orbitals are determined
by a fractionally occupied Hartree--Fock calculation, after which a
(non-self-consistent) radial potential is calculated with the LDA
exchange functional; this scheme will be denoted LDA@HF for the
remainder of the manuscript.  The LDA@HF scheme is an example of
density-corrected DFT,\citep{Kim2014, Vuckovic2019} which sometimes
offers a way to obtain more accurate results in DFT calculations.

As is well known, the electronic structure of atoms is dissimilar from
that of molecules: atoms often exhibit significant spin-polarization,
while molecules are typically singlets;\citep{Perdew2016b} thus, good
performance for atoms does not necessarily imply good behavior for
molecules. As a potential that is both local and scalar is desired for
use in molecular calculations, we consider four ways in which such a
potential can be achieved from an atomic calculation. We study (i)
spin-restricted calculations on the $4s^{2}3d^{6}$ singlet state, as
well as three kinds of potentials from spin-unrestricted calculations
on the $4s^{2}3d^{6}$ quintet state: (ii) the spin-averaged potential,
(iii) the potential from the spin-averaged density, as well as (iv)
the majority-spin potential.

\begin{figure}
\includegraphics[width=0.5\textwidth]{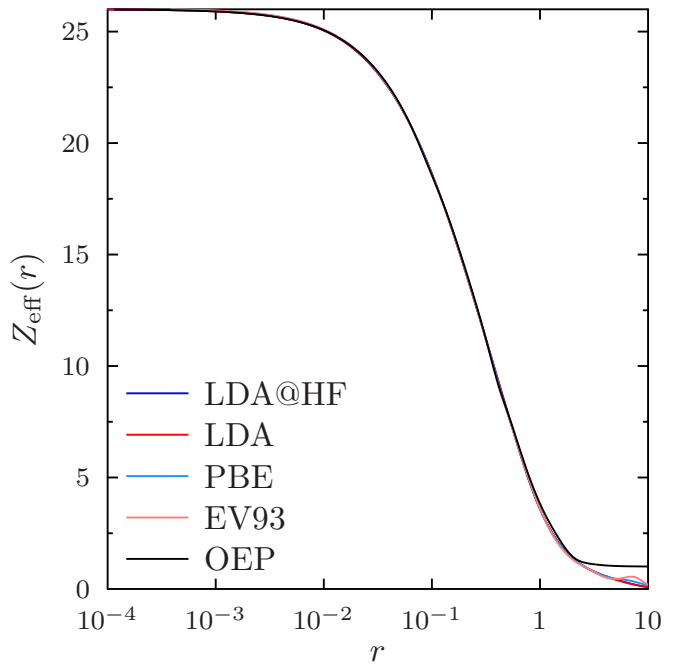}
\caption{Spin-restricted potentials for $4s^{2}3d^{6}$ Fe. The
  potentials are indistinguishable until about $r=a_0$, where the OEP
  starts to approach the asymptotic behavior visible at large
  $r$. \label{fig:Spin-restricted-potentials-for}}
\end{figure}

To begin, we compare the various kinds of density functional
potentials against the OEP for a spin-restricted calculation; the
results are shown in \figref{Spin-restricted-potentials-for}. The
potentials are indistinguishable until about one bohr, where the
density functionals start to diverge from the OEP. The OEP saturates
to its asymptotic limit $V(r)=-1/r$ at a distance of 2 bohr, while the
density functional potentials keep on decaying exponentially.

\begin{figure}
\includegraphics[width=0.5\textwidth]{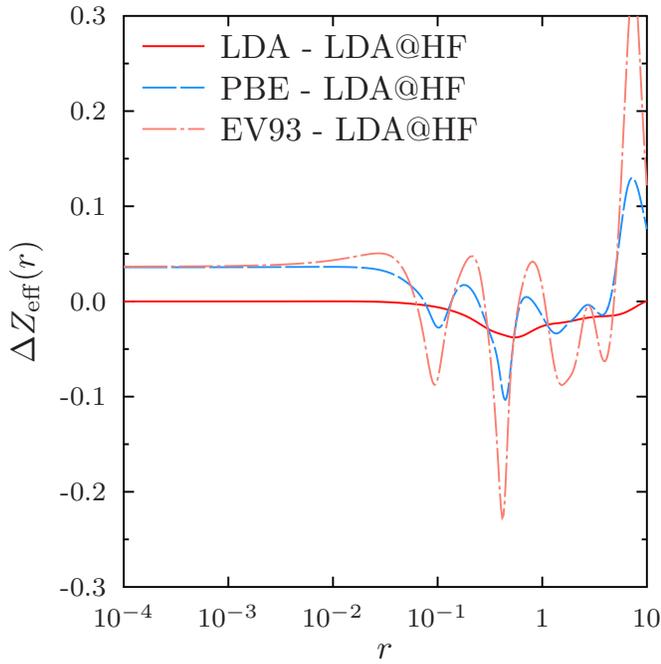}
\caption{Difference of spin-restricted potentials for $4s^{2}3d^{6}$ Fe.\label{fig:Difference-of-spin-restricted}}
\end{figure}

The difference of the self-consistent LDA, PBE and EV93 potentials
from the non-self-consistent LDA@HF potential is further studied in
\figref{Difference-of-spin-restricted}. The EV93 potential clearly has
a lot more structure than LDA@HF, and is characterized by a number of
kinks and sharp peaks. Also the PBE potential is somewhat peaked.  The
LDA and LDA@HF potentials, however, are close to identical --
differring by less than $0.04e$ at any range -- implying that
self-consistency is not that important.

\begin{figure}
\includegraphics[width=0.5\textwidth]{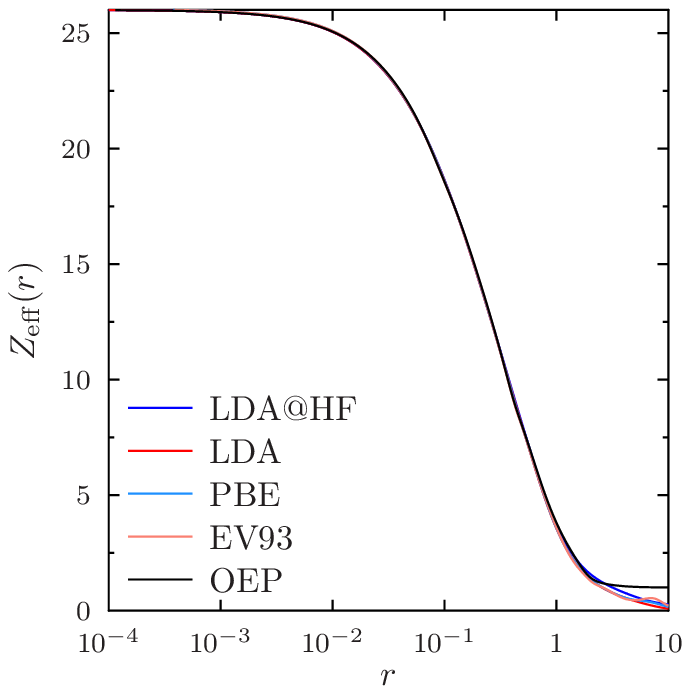}
\caption{Spin-averaged potentials for $4s^{2}3d^{6}$ Fe.\label{fig:Spin-averaged-potentials-for}}
\end{figure}

\begin{figure}
\includegraphics[width=0.5\textwidth]{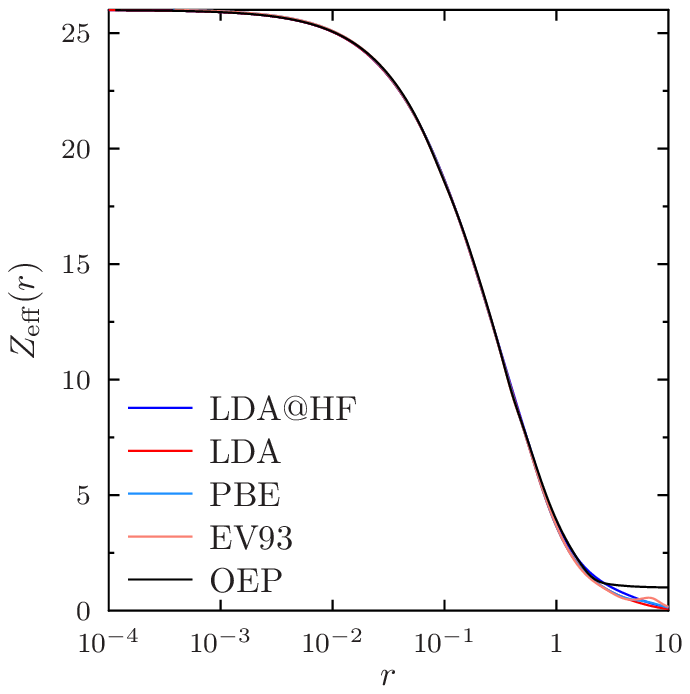}
\caption{Majority-spin potentials for $4s^{2}3d^{6}$ Fe.\label{fig:Spin-majority-potentials-for}}
\end{figure}

Next, the spin-averaged and majority-spin potentials from
spin-unrestricted calculations are shown in
\figref{Spin-averaged-potentials-for, Spin-majority-potentials-for},
respectively, and the differences of the various DFT potentials from
the LDA@HF potential are shown in \figref{Difference-of-spin-averaged,
  Difference-of-spin-majority} for the spin-averaged and majority-spin
potentials, respectively. All three potentials appear surprisingly
similar, from which one can infer that the choice of the density
functional has little effect on the potential for a fixed electronic
configuration.

The use of a fixed electronic configuration is, however, a
restriction: different choices of the functional and of the spin
treatment may yield different ground state configurations. Such
changes are likely to be reflected as larger changes in the potential,
which are not considered here. The optimal electronic configuration to
be used for a starting guess in molecular calculations is probably the
one that is dominant in a molecular environment. For heavy elements
the choice is, however, not straightforward, as they may exhibit many
low-lying states that couple strongly together in a molecular
environment.

\begin{figure}
\includegraphics[width=0.5\textwidth]{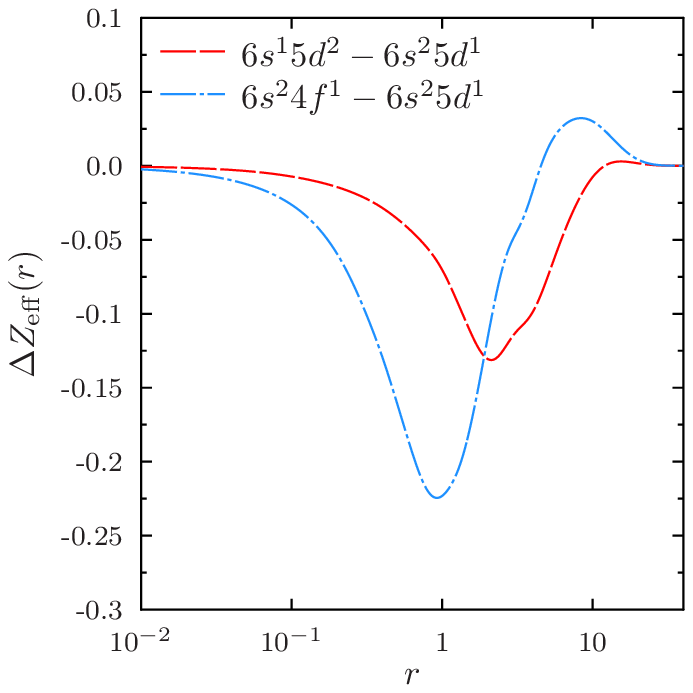}
\caption{Difference of configurations for La computed with LDA@HF at the Dirac--Coulomb level of theory with \Grasp{}.\label{fig:la-configs}}
\end{figure}

To demonstrate the differences in the potential arising from a change
of the reference configuration, in \figref{la-configs} we show the
differences between the LDA@HF potentials computed for the
ground-state $\text{[Xe]}6s^{2}5d^{1}$ configuration, and the
$\text{[Xe]}6s^{1}5d^{2}$ and $\text{[Xe]}6s^{2}4f^{1}$ excited-state
configurations of lanthanum.  These differences are similar in
magnitude to those observed between different functionals for Fe in
$4s^2 3d^6$, suggesting that the choice of the reference configuration
should also be of relatively small importance for the initial guess.

\begin{figure}
\includegraphics[width=0.5\textwidth]{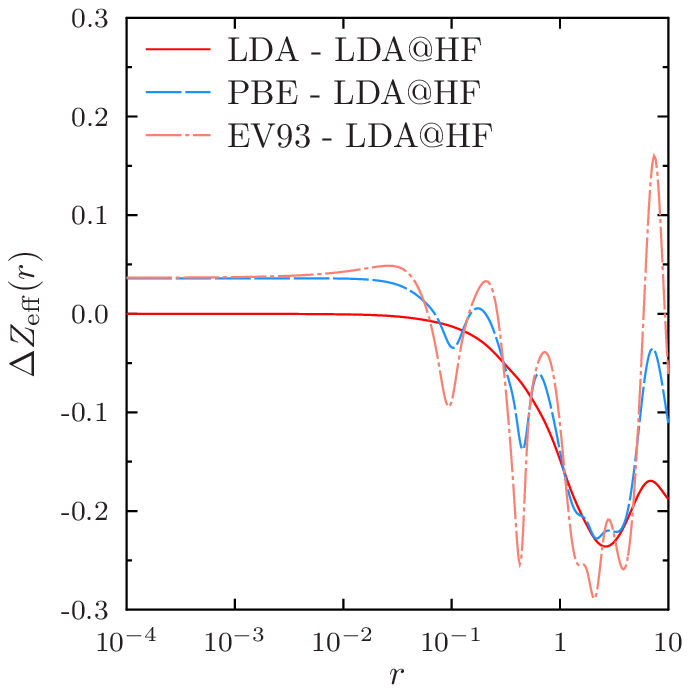}
\caption{Difference of spin-averaged potentials for $4s^{2}3d^{6}$ Fe.\label{fig:Difference-of-spin-averaged}}
\end{figure}

\begin{figure}
\includegraphics[width=0.5\textwidth]{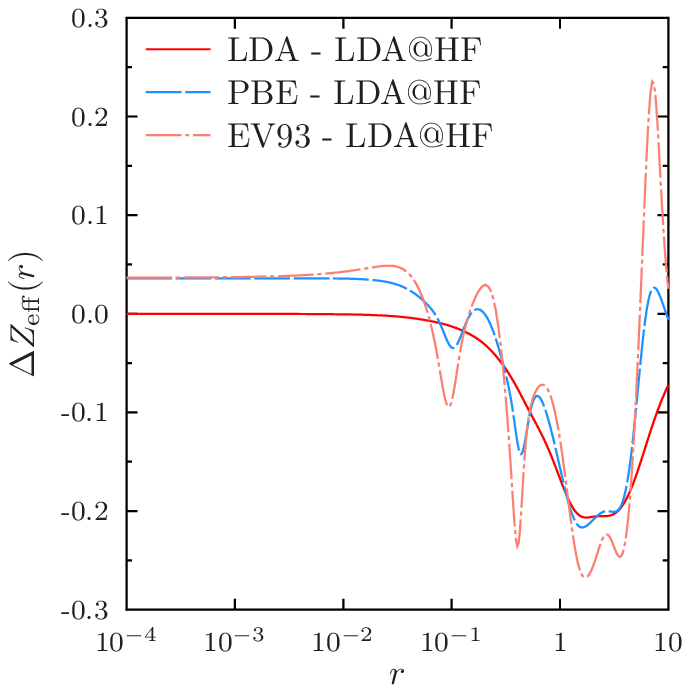}
\caption{Difference of majority-spin potentials for $4s^{2}3d^{6}$ Fe.\label{fig:Difference-of-spin-majority}}
\end{figure}

\subsection{Choice of the potential \label{sec:potchoice}}

We have discussed four different ways in which the effective scalar
potential can be chosen from a (possibly spin-polarized) atomic
calculation, but it is not \emph{a priori} clear which one affords the
best accuracy in molecular calculations due to the characteristic
differences between the electronic structure of atoms and
molecules. As first-row transition metals have significant open-shell
character and spin polarization, and are also infamous for their
several low-lying excited atomic configurations, first-row transition
metal complexes should offer excellent test cases for determining how
the effective atomic radial potential should be formed for molecular
calculations.

We employ the quadrature-based implementation of the SAP guess of
\citeref{Lehtola2019} to assess the four possible choices for the
scalar atomic potential, which were outlined above in
\secref{potform}. Calculations are performed with \Erkale{} on the set
of 32 diverse first-row transition metal complexes of
\citeref{Buhl2006} at the BP86/def2-QZVP\cite{Becke1988a, Perdew1986,
  Weigend2003} level of theory, which yields a good description of the
ground-state geometries.\cite{Buhl2006} Note that this set also formed
part of the test set of \citeref{Lehtola2019}, in which the SAP guess
(based on less accurate radial potentials than in the present work)
was assessed in comparison to several other commonly-used initial
guesses, and was shown to yield good accuracy. A (75,302) grid was
used for the SAP guess and the density functional calculation, and
linear interpolation was used for the tabulated potential. The
geometries of \citeref{Buhl2006} are used, which are optimal for this
level of theory. The universal Coulomb fitting basis set for the def2
series\cite{Weigend2006} is used to reduce the cost of the
calculations.

The \HelFEM{} potentials corresponding to the four choices of the
potential with either LDA or HF orbitals (taken from their
corresponding ground state configurations) yield the results in
\tabref{poterrors}. The best results are obtained with the
spin-averaged potential from unrestricted HF calculations, while the
potential from the spin-averaged density yields the second-best
results. (The unrestricted HF configurations were given above in
\tabref{conf}, while the spin-restricted LDA and HF configurations
were reported in \citeref{Lehtola2019e}.)

\begin{table}
  \begin{tabular}{lr}
    Potential & Mean error ($ E_h $) \tabularnewline
    \hline
    \hline
    LDA@HF (ii)  & 5.679 \\
    LDA (ii) & 7.015 \\
    LDA@HF (iii) & 10.011 \\
    LDA (iii) & 10.232 \\
    LDA (i) & 12.540 \\
    LDA (iv) & 14.400 \\
    LDA@HF (iv) & 14.518 \\
    LDA@HF (i) & 16.428
  \end{tabular}
  \caption{Mean errors in the SAP guess energy for the transition
    metal complex database of \citeref{Buhl2006} at BP86/def2-QZVP
    level of theory for various choices of the LDA or LDA@HF potential
    discussed in \secref{potform}. \label{tab:poterrors}}
\end{table}

\subsection{Error-function fits of the local exchange potential \label{sec:fits}}

Let us summarize the work so far. We found the potentials given by
various functionals to be more or less similar for the $4s^{2}3d^{6}$
states of iron in \secref{potform}, so for simplicity we chose to use
local exchange potentials for the rest of the work. The local exchange
potentials for the various configurations for La were also found to
have similar shapes in \secref{potform}, meaning that any reasonable
choice for the atomic configurations should yield similar results.
Finally, the use of Hartree--Fock orbitals to generate the LDA
potentials was found to result in smaller errors of the guess energy
in \secref{potchoice}, due to which Hartree--Fock orbitals are used
for the rest of this work.

Relativistic potentials are generated with \Grasp{} from
(spin-restricted) average-level Dirac--Coulomb--Hartree--Fock
calculations,\cite{Visscher1997} whereas the non-relativistic
potentials from \HelFEM{} are based on spin-unrestricted Hartree--Fock
calculations,\cite{Lehtola2019e} as multiconfigurational calculations
are not yet available in \HelFEM{}.  In order to have a consistent,
unambiguous and non-empirical choice of configuration, we choose the
configuration with the lowest energy as the reference configuration.
That is, the \Grasp{} calculations generate a potential from the
configuration with the lowest average energy from the AL calculations,
wheras the \HelFEM{} calculations employ the configuration yielding
the lowest energy using a spherically symmetric density
(\tabref{conf}).

\begin{figure}
\includegraphics[width=0.5\textwidth]{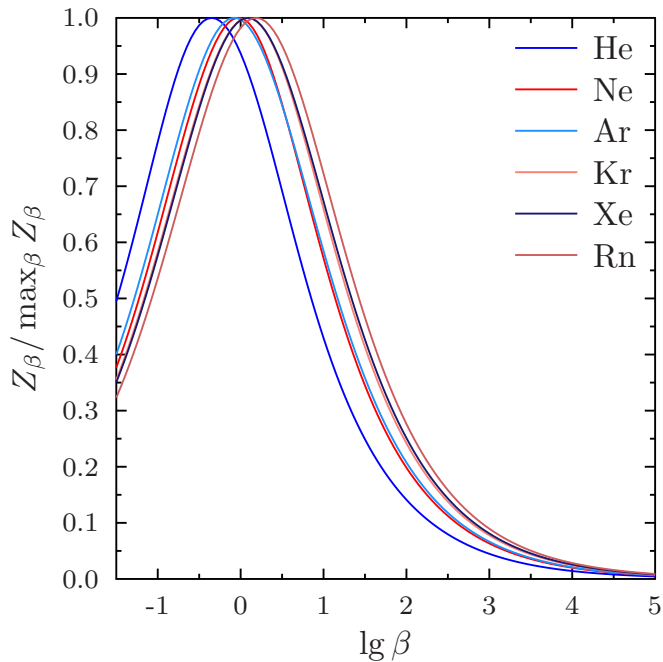}

\caption{Projection of $Z(r)$ onto a normalized complementary error
  function $N(\beta) \erfc{\beta r}$. As these projections still scale
  with $Z$, they have been further normalized to a unit maximum for
  each element.\label{fig:erfc-projection}}
\end{figure}

Before full engagement with the error-function fits, we demonstrate
that the error function basis is suitable for expanding the radial
potential. \Figref{erfc-projection} shows the projections of the
radial potentials for the noble gas series produced by \HelFEM{} onto
normalized complementary error functions according to
\eqref{Zproj}. The similarity of the shape of the projections is
striking: even though the nuclear charge ranges from $Z=2$ for He to
$Z=86$ for Rn, there is but a slight migration to steeper exponents,
with a strong decay for the tight region. Due to the smooth behavior
evidenced by \figref{erfc-projection}, we can start planning the
actual fits.

Exploratory calculations were performed on the alkali and noble gas
atoms with $\beta_0=10^{-2}$ and various choices for $\gamma$ ($1.2$,
$1.3$, \dots, $2.0$), as the alkali metals and the noble gases
represent the most delocalized and the most localized electronic
structure, correspondingly. The calculations (not shown) confirm that
as expected, the fit error decreases monotonically in decreasing
$\beta$, and that suitably accurate fits are achievable with
$\gamma=1.4$; this choice corresponds to a spacing of $\gamma^2 =
1.96$ for the density primitives $\alpha_p$. The choice $\gamma=1.8$
that corresponds to a density primitive spacing of $\gamma^2 = 3.24$
yields a less accurate but computationally cheaper choice for the
fitting basis.

Having fixed the parameters used for the fits, the fits for the whole
set of \HelFEM{} and \Grasp{} data are straightforwardly determined
with the procedure of \secref{fitscheme}.  With the parameters
$\beta_0 = 10^{-2}$ and $\gamma=1.4$, all the \HelFEM{} and \Grasp{}
data are fit by a common set of 26 and 25 parameters, respectively,
where the fit for each element typically uses only a fraction of the
total set of parameters. The maximum fitting errors are
$\tau=1.68\times10^{-3}$ for Dy and $\tau=1.22\times10^{-4}$ for Og
for the \HelFEM{} and \Grasp{} fits, respectively.

Increasing the spacing to $\gamma=1.8$, the \HelFEM{} and \Grasp{}
data are fit by a common set of 17 and 13 parameters, respectively,
each fit again using a subset of the common parameters. The maximum
fitting errors are $\tau=5.68\times10^{-3}$ for Lr and
$\tau=4.27\times10^{-3}$ for No for the \HelFEM{} and \Grasp{}
potentials, respectively. The fitted potentials are available as part
of the supplementary material, in both \Dirac{} and \Gaussian'94 basis
set format.

\section{Molecular applications \label{sec:molecular}}

\subsection{Non-relativistic calculations \label{sec:erkale}}

The accuracy of the fits can be compared to the quadrature results of
\secref{potchoice}; mean errors for error-function potentials as well
as the \Grasp{} quadrature guess are shown in \tabref{comparison}.
(The fitted guess, \eqref{sap-fitpot}, was implemented in \Erkale{}
for this comparison.) The SAP guess by quadrature is by far the most
accurate; however, as the potential was chosen in \secref{potchoice}
to minimize the error for these calculations, this finding may be
somewhat biased. The second-most accurate guess is afforded by the
tabulated potential from \Grasp{}, suggesting that the use of
tabulated, fully numerical potentials as in \citeref{Lehtola2019} is
beneficial for accuracy: the integral on the Becke grid is essentially
exact, which is why DFT calculations are tractable in the first place.

In contrast, the use of a Gaussian fit---whether it is explicitly
optimized for molecules or not---poses limits on the accuracy of the
guess. Interestingly, the errors for the present fits compare
favorably with those for the optimized fits of \citeref{Laikov2020}:
the accuracy of the present unoptimized fits is close to that of the
hand-optimized potentials of \citeref{Laikov2020}. It is also
interesting to note that the mean errors for all of the Gaussian-fit
potentials (in the range of 11 to 16 kcal/mol) are close to those of
the other choices for the fully numerical potential in the quadrature
study in \tabref{poterrors} (10 to 16 kcal/mol).

The database is challenging for all the examined guesses, as errors of
tens of $E_h$ are observed for several molecules for every guess. The
\ce{Sc(acac)3}, \ce{Cu(acac)2}, \ce{Ni(acac)2} complexes (acac =
acetylacetonato) typically belong to the top four worst performers for
all potentials.

\begin{table}
  \begin{tabular}{lr}
    Potential & Mean error ($ E_h $) \tabularnewline
    \hline
    \hline
    LDA@HF (ii), quadrature  & 5.679 \\
    \Grasp{}, quadrature & 9.785 \\
    optimized HF fit$^{a,b}$  & 11.531 \\
    LDA@HF (ii), large fit  & 11.743 \\
    LDA@HF (ii), small fit  & 11.836 \\
    optimized LDA fit$^a$  & 12.244 \\
    \Grasp{}, large fit & 15.879 \\
    \Grasp{}, small fit & 16.699
  \end{tabular}
  \caption{Mean errors in the SAP guess energy for the transition
    metal complex database of \citeref{Buhl2006} at BP86/def2-QZVP
    level of theory. $^a$Fits from \citeref{Laikov2020}. $^b$The HF
    potential was capped to charge neutrality. \label{tab:comparison}}
\end{table}

\subsection{Relativistic calculations \label{sec:dirac}}
The fits of the \Grasp{} potential derived in this work differ from
the unpublished fits of \Dirac 16 which were similarly based on
LDA@HF, but also included a correlation contribution in the potential
from the Vosko--Wilk--Nusair functional.\cite{Vosko1980} Moreover, an
accurate, but computationally unoptimized fit to 30 functions with
$\beta_0 = 10^{-2}$ and $\gamma=\sqrt{2}$ was used in \Dirac 16. The
fits of the present work are both more economical in terms of number
of functions, as well as more accurate due to the exact adherence to
the sum rule of \eqref{c-cond} and the use of the complementary error
function basis with analytical overlap. Although the present fits are
smaller and do not include a correlation potential, the number of
iterations needed to converge Hartree--Fock calculations remains
similar, illustrating again the minor importance of the correlation
functional for the starting potential as well as the small effect on
self-consistent field convergence arising from minor changes to the
atomic potential.

As an example, we show three illustrative cases for the performance of
the SAP starting potentials. The \ce{K2} dimer with an internuclear
separation of 4.0 Å probes the long-range part of the potential, CsCl
with a bond distance of\cite{NIST-webbook} 2.906 Å represents a
prototypical ionic bond, and the octahedral \ce{UF6} molecule with an
\ce{U\bond{-}F} distance of\cite{Kimura1968} 1.996 Å tends to converge
onto a higher-lying SCF solution with other starting procedures. The
all-electron Dirac--Coulomb--Hartree--Fock calculations used the
triple zeta basis sets developed by Dyall,\cite{Dyall2006a, Dyall2009,
  Dyall2016a} with the default approximation of neglecting the (SS|SS)
small-component Coulomb integrals.\cite{Visscher1997a} All
calculations with the SAP guess converge smoothly within 13 (\ce{K2}
and CsCl) or 18 (\ce{UF6}) iterations. As a further measure for the
goodness of the guess, in \tabref{molecular_energies} we compare the
converged Dirac--Coulomb--Hartree--Fock energy to that of the first
iteration, which is based on the orbitals that result from the
diagonalization of the summed atomic potentials. The relatively small
energy differences demonstrate the adequacy of even the small fit in
the core and valence regions of the heavy atoms.

\begin{table}
\caption{Converged Dirac--Coulomb--Hartree--Fock energy $E^\text{final}$, and error in
  the large and small fit SAP guess energy $\Delta E^\text{SAP} = E^\text{SAP} -
  E^\text{final}$ for \ce{K2}, CsCl, and \ce{UF6}.}
\begin{center}
\begin{tabular}{crrr}
Molecule & $E^\text{final}$ ($E_h$)  & $\Delta E_\text{large}^\text{SAP}$ ($E_h$) & $\Delta E_\text{small}^\text{SAP}$ ($E_h$) \\
\hline
\hline
\ce{K2}  &  $-1203.033795$ & $0.075046$ & $0.070591$ \\
CsCl     &  $-8248.076864$ & $0.145937$ & $0.189300$ \\
\ce{UF6} & $-28656.637434$ & $0.438916$ & $0.593749$
\end{tabular}
\end{center}
\label{tab:molecular_energies}
\end{table}

\section{Summary and Discussion \label{sec:summary}}

The superposition of atomic potentials (SAP) guess\cite{Lehtola2019}
builds guess orbitals for electronic structure calculations from a
simple sum of atomic effective potentials. We have compared atomic
effective potentials from density functional and optimized effective
potential calculations, and found their differences to be relatively
small if the configuration is fixed. Our results suggest that atomic
local density exchange potentials should offer a good starting point
for electronic structure calculations. We have also compared the local
density exchange potentials arising from different choices for the
atomic configuration, and found also these differences to be
small. Due to this, it is our belief that any reasonable choice for
the reference atomic configuration should yield suitable starting
guesses; minimal-energy Hartree--Fock configurations were used in the
present work.

We have pointed out that the SAP guess can be easily implemented in
Gaussian-basis quantum chemistry programs by fitting the fully
numerical, complete-basis-set-limit radial effective potentials in
terms of error functions. We have described a robust method for
forming such fits, and reported two sets of fits at two levels of
accuracy available in the supplementary material, consisting of a
common set of 17 and 26 $s$-type primitives, respectively, which are
suitable for inclusion in electronic structure programs. Fits were
formed both at the non-relativistic and fully relativistic levels of
theory to suit the needs of all applications.  As the fits consist of
just one highly contracted $s$ function, the computation of the
resulting matrix elements as three-center two-electron integrals is
extremely rapid even in large orbital basis sets.

The most commonly used initial guess nowadays is the superposition of
atomic densities (SAD),\citep{Almlof1982, VanLenthe2006} in which a a
molecular calculation is initialized by a block-diagonal density
matrix arising from a set of atomic calculations. The SAP guess is
more aesthetically pleasing than SAD: SAP yields guess orbitals
straight away, whereas SAD requires a full Fock matrix build from the
non-idempotent guess density matrix, which can then be diagonalized to
yield orbitals. In SAP, in contrast, the guess orbitals are obtained
by diagonalizing the approximate one-electron Hamiltonian, in which
the molecular potential is estimated directly as a superposition of
pretabulated atomic potentials. The matrix elements necessary for SAP
can be easily implemented by quadrature as in \citeref{Lehtola2019};
alternatively, as discussed in the present work, error-function fits
to the atomic potentials allow reformulating the guess in terms of
two-electron integrals in Gaussian-basis programs.

Despite their significant formal difference, the SAD and SAP
approaches are quite similar at the complete basis set limit. In
either case, the Coulomb part of the potential will be the same, as it
is linear in the density: the Coulomb potential arising from the sum
of spherically symmetric atomic densities is the same as the sum of
spherically symmetric atomic Coulomb potentials. However, the
situation is trickier for the exchange. Although the exact exchange
operator is linear in the density matrix, the corresponding local
scalar potential may behave discontinuously in the number of
electrons.\cite{Hodgson2017} Because of this, the approaches can be
better contrasted within a density functional approximation: SAD
yields a local exchange potential
\begin{equation}
  V_x^\text{SAD}({\bf r}) \propto - \left[ \sum_A n_A({\bf r}) \right]^{1/3} \label{eq:Vsad}
\end{equation}
whereas SAP yields
\begin{equation}
  V_x^\text{SAP}({\bf r}) \propto - \sum_A \left[n_A({\bf r})\right]^{1/3}. \label{eq:Vsap}
\end{equation}
This appears to suggest that SAP is more attractive than SAD in
molecules, meaning that SAD and SAP can always be expected to
reproduce different results---even at the complete basis set
limit. (Note that since generalized gradient approximation (GGA) and
meta-GGA functionals build on top of the local exchange functional, a
similar argument should also hold for them.) Although in many cases
both SAD and SAP will lead to rapid convergence to the same
ground-state in a self-consistent field calculation, systems with
challenging electronic structures may exhibit several physical
solutions of different symmetry or charge localization. Access to
different types of guesses is extremely helpful in cases where a
single guess does not perform adequately. Although several programs
offer simple choices to SAD, such as the core Hamiltonian
a.k.a. one-electron guess, these choices may be of extremely poor
accuracy for large systems in contrast to SAD and SAP that both
account for screening effects in heavy atoms.\cite{Lehtola2019} The
present fitted atomic potentials facilitate the inclusion of the SAP
guess in commonly-used Gaussian-basis quantum chemistry programs,
thereby introducing a new class of accurate initial guesses that may
aid studies on challenging systems.

\section*{Supplementary material}

See supplementary material for the small ($\gamma=1.8$) and large
($\gamma=1.4$) fits to the \HelFEM{} and \Grasp{} potentials, in both
\Dirac{} and \Gaussian'94 format.

\section*{Acknowledgments}

This work has been supported by the Academy of Finland (Suomen
Akatemia) through project number 311149. Computational resources
provided by CSC -- It Center for Science Ltd (Espoo, Finland) and the
Finnish Grid and Cloud Infrastructure (persistent identifier
urn:nbn:fi:research-infras-2016072533) are gratefully acknowledged.

\bibliography{citations, citations_luuk}

\end{document}